\documentclass{article}
\usepackage{amsmath, amssymb}
\usepackage{geometry}
\usepackage{graphicx}
\usepackage{caption} 
\usepackage{authblk}
\usepackage[numbers,square,sort&compress]{natbib}
\usepackage{booktabs}
\usepackage{url}
\usepackage[colorlinks]{hyperref} 
\usepackage{comment}
\usepackage{color}

\title{Comment on ``Average Hazard as Harmonic Mean'' by Chiba (2025)}

 \author[1,2]{Hajime Uno\thanks{Corresponding author: huno@ds.dfci.harvard.edu}}
 \author[3]{Lu Tian}
  \author[1,4]{Miki Horiguchi}
  \author[5]{Satoshi Hattori}

  \affil[1]{Department of Data Science, Dana-Farber Cancer Institute, Boston, MA, USA}
  \affil[2]{Department of Medicine, Harvard Medical School, Boston, MA, USA}
  \affil[3]{Department of Biomedical Data Science, Stanford University, Stanford, CA, USA}
  \affil[4]{Department of Biostatistics, Harvard T H Chan School of Public Health, Boston, MA, USA}
  \affil[5]{Department of Biomedical Statistics, Graduate School of Medicine and Integrated Frontier Research for Medical Science Division, Osaka University, Osaka, Japan.}

\date{}

\begin{document}
\maketitle

\section*{Introduction}
For event time $T,$ the average hazard \cite{Uno2023-sm} (or generalized hazard \cite{Snapinn2022-wf,Jackson2025-vq}) over a given time range $[0,\tau]$ is defined as  
\begin{equation*}
AH(\tau) = 
\frac{E\{ I(T\le \tau) \}}{E\{ \min(T,\tau)\}} = 
\frac{1-S(\tau)}{\int_0^{\tau} S(u) du},
\label{AH}
\end{equation*}
where $I(A)$ is an index function that takes 1 if $A$ is true and 0 otherwise, 
and $S(t)=\Pr(T>t)$ is the survival function for $T.$ 
Note that the numerator, $E\{ I(T \le \tau) \}, $ is the cumulative incidence probability at $\tau$, and the denominator, $E\{ \min(T,\tau)\},$ is the restricted mean survival time (RMST) with the truncation time $\tau.$   
The Kaplan-Meier method is a well-established nonparametric method to estimate the survival function and is commonly used in various survival data analyses. Similar to the cases estimating the cumulative incidence probability and RMST,  
$ AH(\tau) $ can also be nonparametrically estimated by plugging in the Kaplan-Meier estimator $\hat{S}(\cdot)$ for the $S(\cdot)$ as follows,  
\begin{equation}
\widehat{AH}(\tau) = 
\frac{1-\hat{S}(\tau)}{\int_0^{\tau} \hat{S}(u) du}.
\label{AHEST}
\end{equation}
Asymptotic properties, such as consistency, asymptotic normality, and asymptotic variance have already been reported, along with the finite sample performance of this estimator based on simulation studies, such as bias and coverage probabilities of the corresponding confidence intervals \cite{Uno2023-sm}. 

In a recent article published in Pharmaceutical Statistics, {\it ``Average Hazard as Harmonic Mean''} (Chiba, 2025) \cite{Chiba2025}, the author re-interprets the average hazard as the harmonic mean of the hazard function (Section 2) 
$$ AH(\tau) = 
\frac{\int_0^\tau f(t)dt}{\int_0^{\tau} f(t)/h(t) dt},$$ 
where $f(t)$ and $h(t)$ are the density function and hazard function, respectively. 
Chiba (2025) then asserts, in Section 3, that the Kaplan–Meier plug-in estimator (\ref{AHEST}) is \emph{``incorrect''} whenever the truncation time $\tau$ is not an {\it observed} event time.

In this note we demonstrate that the argument in Chiba (2025) is challenged by several conceptual and methodological gaps.  
By disentangling these points, we reaffirm that the Kaplan–Meier plug-in estimator~(\ref{AHEST}) remains sound for estimating $AH(\tau)$—and, above all, we show that investigators can continue to apply it with confidence.

\section*{Logical Gaps in Chiba's Proof by Contradiction (Chiba Section 3.2)}

Before analyzing the proof itself, we first note a foundational gap: Chiba (2025) did not define what it would mean for the Kaplan–Meier plug-in estimator (\ref{AHEST}) labeled as ``Formula (4)'' to be {\it correct}.  
The manuscript does not clarify whether ``correctness'' of the estimator should be interpreted as unbiasedness, consistency, or some other statistical property.  Because this premise remains undefined, the logical footing of the forthcoming proof-by-contradiction is weak: one cannot rigorously prove or disprove a vague statement. 

Let \(t_{1} < t_{2} < \dots < t_{K}\) denote the distinct {\it ``observed''} failure times, listed in ascending order. 
The Kaplan–Meier plug-in estimator (\ref{AHEST}) is then rewritten as
\begin{equation*}
\widehat{AH}(\tau)=
\frac{1-\hat S(\tau)}
     {\sum_{j=1}^{i}\hat S(t_{j-1})(t_j-t_{j-1})+\hat S(t_i)(\tau-t_i)},
     \qquad t_i<\tau<t_{i+1},
\end{equation*}
which is referred to as ``Formula (4)'' in Chiba's paper. In this note, we also call this Formula (4) thereafter.
Chiba (2025) rewrites the Formula (4) into their ``harmonic–mean'' form as follows, 
\begin{equation*}
\frac{\displaystyle\sum_{j=1}^{i}\hat f(t_j)(t_j-t_{j-1})+\hat f(\tau)(\tau-t_i)}
     {\displaystyle\sum_{j=1}^{i}\dfrac{\hat f(t_j)(t_j-t_{j-1})}{\hat h(t_j)}+
       \dfrac{\hat f(\tau)(\tau-t_i)}{\hat h(\tau)}},
\label{eq:4prime}
\end{equation*}
where 
$\hat f(t_j)=\hat h(t_j)\,\hat S(t_{j-1}),$  
$\hat h(t_j)=d_j/[r_j(t_j-t_{j-1})],$  
and $d_j$ and $r_j$ denote the numbers of observed events and subjects at risk at time \(t_j\), respectively.
Chiba (2025) then claims in their proof-by-contradiction that  
{\it ``Under the assumption that Formula (4) is correct for $t_i < \tau < t_{i+1},$
this formula for $\widehat{AH}(\tau)$ implies that
$\hat{f}(\tau) = 0$ and
$\hat{f}(\tau)/\hat{h}(\tau) >0 $ 
because 
$\hat{S}(\tau) = \hat{S}(t_i)$ 
and
$\hat{S}(t_i) (\tau - t_i) > 0$ in Formula (4).
However, it is obvious that
$\hat{f}(\tau) = 0$ 
contradicts 
$\hat{f}(\tau)/\hat{h}(\tau) >0.$ 
Therefore, the assumption is false; that is, `Formula (4)' is incorrect for $t_i < \tau < t_{i+1}.$'' 
}

In short, Chiba asserts that $\hat f(\tau)=0$ conflicts with $\hat f(\tau)/\hat h(\tau)>0$, thereby declaring Formula (4) ``incorrect.'' 
Yet no real contradiction is established here, because the ratio $\hat f(\tau)/\hat h(\tau)$ becomes the indeterminate form $0/0,$ when $\hat h(\tau)=0.$
Chiba also acknowledges that {\it ``$\hat h(t_j)$ cannot be defined at a time at which the event is not observed; thus, $\hat{h}(\tau)$ cannot be defined.''} (p.\,3, col.\,2, l.\,9–10), which suggests that neither $\hat h(\tau)$ nor $\hat{f}(\tau)/\hat{h}(\tau)$ is well defined for $t_i<\tau<t_{i+1}.$ 
Consequently, the alleged contradiction in Section 3.2 doesn't exist.

\section*{Sampling Variability Misinterpreted (Chiba Section 3.1) }

In Section 3.1, Chiba uses a sample data set
$$ 10, 21, 34, 48, 65, 85, 109, 120^{*}, 120^{*}, 120^{*} (days) $$
and states,  {\it ``Formula (4) may not estimate the average hazard appropriately when $t_i<\tau<t_{i+1}.$''} 
Chiba first draws the cumulative hazard plot using the Nelson-Aalen estimator (Figure 1 in Chiba) and says that the hazard appears ``constant.''
Second, Chiba calculates the $AH(\tau)$ with various $\tau$ values using Formula (4) (Figure 2 in Chiba). 
Lastly, Chiba comments {\it ``as shown in Figure 2, the values of $\widehat{AH}(\tau)$ are not constant for $\tau \neq t_i,$ while those are approximately constant for $\tau = t_i.$ This observation shows that the average hazard may not be estimated appropriately when the truncation time is set to a time at which the event is not observed.''}

Here, we summarize several issues on reasoning Chiba used.

\subsubsection*{The underlying distribution is not clear}
Chiba never clearly states the underlying distribution from which their sample data are generated. Therefore, the true cumulative hazard function or true average hazard is unknown. Without knowing the true population parameter, it is impossible to determine if an estimator estimates the true parameter value well.

\subsubsection*{Unbalanced judgment between the Nelson-Aalen and the average hazard estimator}
Suppose that Chiba's sample data are from an exponential distribution (i.e., a constant hazard model). In this case, the finite-sample deviation Chiba is concerned with is seen in estimating not only the average hazard (Figure 2 in Chiba) but also the cumulative hazard function (Figure 1 in Chiba). Thus, the Nelson-Aalen estimator also fails to appropriately estimate the cumulative hazard function because it is not a straight line. It obviously contradicts the well-established results on the validity of Nelson-Aalen estimator.

\subsubsection*{Sampling variability should be taken into account} 
When discussing the validity of an estimator in finite sample cases, 
standard practice in statistical research is to evaluate its performance through simulation studies with repeated sampling.  Specifically, to examine the validity of Formula (4) for estimating $AH(\tau)$, we conduct the following simulation study.

We generate the event times randomly from the exponential distribution with a constant hazard of 0.01. To mimic Chiba's sample data, we censor the event time at 120. For each set of simulated data of size $n \in \{10, 30, 50, 100\}$, we estimate the average hazard based on Formula (4) with a range of $\tau$'s. We repeat this process 1000 times and calculate the average of the average hazard estimates and compare them with the true average hazard of 0.01. Figure A shows the results with different sample sizes.  

\begin{figure}[h]
\label{figureA}
\centering
\includegraphics[scale=0.8]{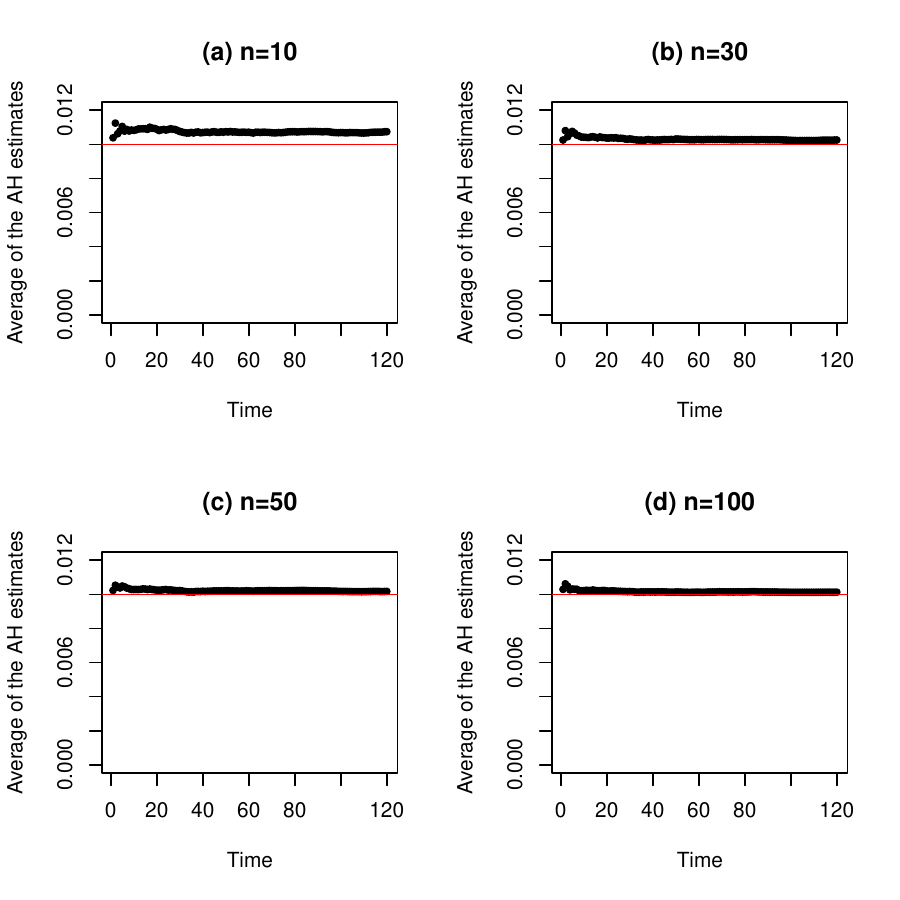}
\caption*{Figure A: Average deviation from the true average hazard (red solid line).}
\end{figure}

Even with a sample size as small as 10 (Figure A, panel (a)), the finite-sample bias of the estimator for the average hazard given by Formula (4) remains negligible for any $\tau\le 120.$ This finding shows that the concern raised by Chiba disappears after averaging over multiple simulation replications.

\section*{Continuous- versus Discrete-Time Concepts}

Chiba (2025) never states the distribution assumption in their argument, continuous time or discrete time. 
In Section 2, it seems that Chiba adopts the continuous-time framework, introducing the usual density function $f(t)$, hazard function $h(t)$, and expressing the average hazard with $f(t)$ and $h(t).$
However, the discussion in Section 3 shifts to a discrete-time view: to support the argument that $\hat h(t)$ is only defined at observed event times, Chiba states that {\it ``$\hat h(t_j)$ cannot be defined at a time at which the event is not observed.''} (p. 3, col. 2, l. 9–10)

Criticizing the validity of an estimator designed for continuous-time data in analyzing discrete-time data is not appropriate, otherwise standard methods, such as the Nelson–Aalen and Kaplan–Meier estimators, would also be ``incorrect'', since they can be expressed in terms of Chiba's $\hat{h}(t)$ as in Formula (4), which is not well defined beyond observed event times.

\subsection*{Why $\widehat{AH}(\tau)$ Based on Formula (4) Is Not Flat Between Events}
Chiba states that {\it ``the average hazard should be estimated using only the times at which the event is observed.''} (p.\,3, col.\,1, l.\,4 from the bottom), which implies that $\widehat{AH}(\tau)$ should remain flat on every open interval $(t_i,t_{i+1})$, just as the Kaplan–Meier and Nelson–Aalen estimators do. Indeed, Kaplan–Meier and Nelson–Aalen curves are step functions: they ``carry forward'' the last value because nothing happens between $t_i$ and $t_{i+1}$.  
The average hazard is different: while its numerator
  $1-\hat S(t)$ stays flat, its denominator
$\int_0^{t}\hat S(u)\,du$ keeps accumulating person‐time, so
$\widehat{AH}(t)$ declines on $(t_i,t_{i+1})$.  The following case scenario will help understand why this makes sense.

\paragraph{A concrete scenario: $S(t)$ is flat.}
Consider an explicit data-generating mechanism in which the true survival curve becomes \emph{exactly horizontal} between two specific time points. 
\begin{center}
\begin{tabular}{lccl}
\toprule
Time interval & Hazard $h(t)$ & Survival $S(t)$ & Comment \\  
\midrule
$0 \le t < 2$  & $1$  & $\exp(-t)$  & {Events will be observed}\\
$2 \le t < 5$  & $0$  & $\exp(-2)$  & {Survival stays flat (no events)}\\ 
$t \ge  5$      & $1$  & $\exp(-t+3)$  & {Events will be observed}\\
\bottomrule
\end{tabular}
\end{center}
Because $S(t)$ is constant for $2\le t \le 5$, the numerator of $AH(t),$
$1-S(t)=1-\exp(-2)$ freezes, whereas the denominator
$\int_0^{t}S(u)\,du$ grows linearly with~$t$.  Hence
\[
  AH(2)=\frac{1-\exp(-2)}{\int_0^{2} \exp(-u) du}=1,
  \qquad
  AH(5)=\frac{1-\exp(-2)}{\int_0^{2} \exp(-u) du + \int_2^{5} \exp(-2) du }
  \approx0.68<AH(2).
\]
Nothing happens between $2$ and $5$ in terms of new events, yet the average hazard drops by about one-third.  Hence a flat survival curve does \emph{not} imply a flat average hazard; the accumulation of person-time without additional failures pushes $AH(t)$ downward (Figure B). 
The same reasoning applies to estimators: on intervals where the Kaplan–Meier and Nelson–Aalen curves are flat, $\widehat{AH}(t)$ based on Formula (4) 
should, as expected, decline---exactly as observed in Chiba's Figures 1 and 2.  

\begin{figure}[h]
\label{figureB}
\centering
\includegraphics[scale=0.7]{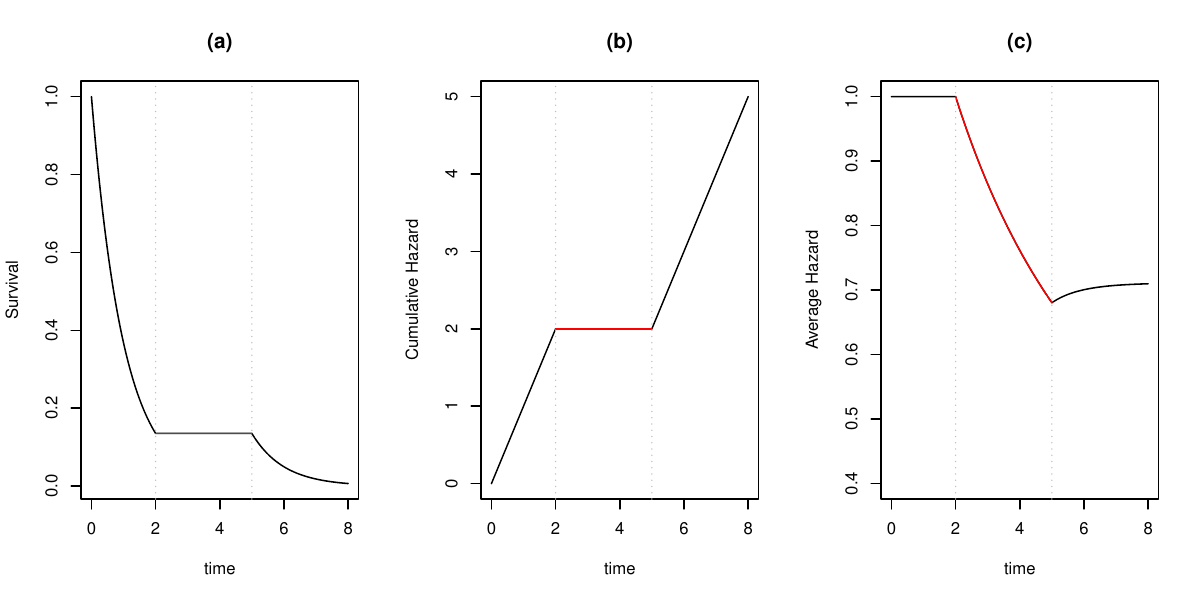}
\caption*{Figure B: (a) Survival function, (b) Cumulative hazard function and (c) Average hazard under a scenario where survival function and cumulative hazard function are flat between 2 and 5.}
\end{figure}

\section*{Conclusion}

Chiba’s critique rests on two unsupported premises: an algebraic rewrite that breaks down once the undefined or zero value of $\hat h(\tau)$ is acknowledged, and a single-sample illustration that confuses finite-sample variation with systematic bias.  When sampling variability is taken into account, the Kaplan–Meier plug-in estimator~(\ref{AHEST}) remains approximately unbiased regardless of whether $\tau$ falls between observed event times, as our simulation study demonstrates. 

In summary, Formula (4) is not ``incorrect''; rather, the contradiction claimed in Chiba (2025) results from applying discrete-time logic to a continuous-time estimator.  We hope this clarification dispels unwarranted skepticism and reinforces confidence in the Kaplan–Meier plug-in estimator for the average hazard. An R implementation of the method is available in the \texttt{survAH} package on the CRAN website\cite{Uno2023-an} and GitHub (\url{https://www.uno1lab.com/survAH/})\cite{survAHrepo}.

\subsection*{Data availability statement}
The data that support the findings of this study are available from the corresponding author upon reasonable request.

\subsection*{Conflict of interest}
The authors declare no conflict of interest.

\subsection*{Funding Information}
This work was supported by the National Institute of General Medical Sciences of the National Institutes of Health under award number R01GM152499 (HU, LT), and the National Heart Lung and Blood Institute of the National Institutes of Health under award number R01HL089778 (LT).

\bibliographystyle{WileyNJD-AMA}
\bibliography{refs_ah} 

\end{document}